# Caring Trouble and Musical AI: Considerations towards a Feminist Musical AI

**Kelsey Cotton[1] Kıvanç Tatar[1]**

[1]**Interaction Design, Chalmers University of Technology**

**ABSTRACT**

The ethics of AI as both material and medium for interaction remains in murky waters within the context of musical and artistic practice. The interdisciplinarity of the field is revealing matters of concern and care, which necessitate interdisciplinary methodologies for evaluation to trouble and critique the inheritance of 'residue-laden' AI-tools in musical applications. Seeking to unsettle these murky waters, this paper critically examines the example of *Holly+*, a deep neural network that generates raw audio in the likeness of its creator Holly Herndon. Drawing from theoretical concerns and considerations from speculative feminism and care ethics, we care-fully trouble the structures, frameworks and assumptions that oscillate within and around *Holly+*. We contribute with several considerations and contemplate future directions for integrating speculative feminism and care into musical-AI agent and system design, derived from our critical feminist examination.

# Introduction

Growing concerns of algorithmic bias and oppression [1] [2] [3] [4] [5]; dataset ownership and data access [6]; and general lore [7] around what AI troubles in our capitalist society is of increasing concern within popular discourse [8] [9]. We see this as an urgent area of concern within musical applications and contexts, which see the integration and assimilation of 'residue-laden' AI systems into musical praxis, musical artworks, and even as synonymous with prominent practitioners.

Within the field of musical-AI, current discourse has examined development of novel tools and architectures for creation and performance [10], artistic potentials in friction and fallacy [11], and human-AI musical interaction [12][13]. Our motivation in this paper is to address longitudinal concerns around the embedding of values, and implications for musical futures. This has previously been alluded to in the literature [14], and we bring a broadening shift towards how AI technologies are changing the landscape of musical creativity [15]. Existing work [16] in evaluating and critiquing AI technologies deployed in performance and artwork contexts has argued for context-specific approaches to system evaluation, but with little exploration into inter-contextual framings of the technology. Emerging concerns regarding AI's influence in curating and shifting musical culture have been outlined in [15] [17], with propositions for policy interventions and the need for future research to examine alternative economic models, longitudinal study and greater diversity.

In this paper, our approach to addressing the myriad of concerns is to examine novel approaches of analysis, which contribute to the field by revealing pathways for workable and ethical practices in the design of musical-AI systems. To do so, our critical examination therefore requires the inclusion of knowledge from other disciplines [18]. We therefore draw together perspectives and methodologies from AI-ethics, Human-Computer Interaction (HCI), science and technology studies (STS), and feminism. The intention with such interdisciplinarity is to uncover and to situate the 'matters of concern' [19] particular to (and within) the field, as evident in our evaluation of our case study. We see interdisciplinarity in approaches to musical-AI as vitally necessary for the community to consider the implications of AI-artworks put out into wider society.

It's now time to look at what's beneath the murky surface of Musical-AI, and to unsettle the water. To assist in our exploration of how concerns are echo-ed between STS, HCI, and care ethics, we care-fully[1] trouble dimensions of Holly Herndon's artwork *Holly+*, a deep neural network that generates audio reminiscent of Herndon's unique vocal aesthetic. We have chosen *Holly+* as a precursory example of how an artist has navigated popular media discourses; collaborative approaches to identity concerns; and articulations of self-governance in the construction and presentation of the artwork. We see our care-full troubling of *Holly+* as an example in taking a step forward from the initial discussions we posed in [20]. In [21] we advocated and argued for conversations on data to expand beyond a generalist and larger-society-centric viewpoint, to elicit domain-specific conversations in the field of musical AI. *Holly+* is both a starting point for the conversations that need to be had, and an example of an artwork that is actively excavating these issues and walking across the borders.

Our contributions here are multifold. First, we contribute with an example of an interdisciplinary analysis, drawing from 3 methodologies in STS, HCI and AI ethics: speculative feminism [22][23]; matters of concern [19] and care [24]; and feminist data ethics [25]. Second, our interdisciplinary analysis reveals 'matters of concern' [19] in the case study, which we argue have implications on the musical-AI community in issues of data, management and legacy. This contribution is augmented by our inclusion of a Knowledge Map [26], to help visibilise the 'matters of concern' and the potential connections between concerns. Third, we outline considerations and future directions for embedding speculative feminism and care into musical-AI design. The stance that we occupy is a *first* step towards provoking change within our community, guided by our concerns as designers, artists, developers and users of the selfsame systems and technologies we critique.

To provide a structural outline of this paper: in the forthcoming Background section, we provide a brief summary of theoretical perspectives and core concepts critical to this paper's inquiry. Chief amongst these are: compounding stances of fact, concern and care; a glimpse into speculative feminist perspectives in STS and HCI; and an overview of current practices in feminist AI-ethics. Drawing upon this theoretical grounding, we then progress to our critical examination of *Holly+*. Extrapolating from our critique, we close by outlining important considerations for inviting speculative feminism into wider discourses on musical-AI and speculate on future directions.

## Background

This paper draws heavily from a breadth of theoretical intersections and disciplines—forming the interdisciplinary foundation of our analysis—and which we will take a moment to address now.

## Matters of Fact, Matters of Concern and Matters of Care

As we examine various dimensions within *Holly+*, Bruno Latour's notion of 'matters of fact' and 'matters of concern' provide assistive concepts [19]. Latour establishes a relation between fact and concern as an act of positioning the objective in relation to the "whole scenography" of its contextual environment. When we consider an AI-agent or AI-system structure as further constituting the objective (matters of fact) in relation to wider contexts (matters of concern), this necessitates a deliberate and care-full troubling [19]. de la Bellacasa unsettles matters of fact and matters of concern [27] with 'matters of care' [24]. Care is defined as engaging with the *becoming* of matters of fact and concern: an intentional *seeking out* of the histories and values present in systems. It is through *seeking out* the histories and contextual entanglements of technologies that *care* is enacted. We see the contribution of these theories as assistive in attending to the positioning of *Holly+* in relation to its intersecting contextual environments, and its becoming. We attend to one of these intersecting environments in our following discussion of feminism in STS.

## Feminism in STS

Looking towards the *what* of what feminism *is*, this helps to formulate the *why* and to establish what feminism can do for musical AI in addressing the inequities, questionable practices and working cultures that are being developed within the music-technology community. This is especially pertinent to musical-AI, which often inherits and adopts models, algorithms and approaches which may carry residues of inequity and bias. Feminist principles can be of benefit here. Speaking broadly, feminism is a series of socio-political movements [28] [29] [30] which seek to address systems of oppression within society, which can encompass gender, political, economic, personal, and social inequalities [31] [32] [33] [34] [35]. Notable work in STS in this regard is the work of Donna Haraway [22] [23] [36], offering speculative feminist narratives of socio-cultural co-construction and fusion of synthetic and organic bodies. Of specific relevance to this paper is their Implosion Analytical Method [37] [38], which critically examines various dimensions of an artefact.

## Feminist Stances and Methodologies in AI and Data Ethics

Looking into broader communities, there is ongoing discourse into the formation and implementation of feminist perspectives [39] [40] [41] in data ethics. Specifically, we will now draw attention to Carroll et al [42], and Gray and Witt [25], who examine critical concerns pertaining to issues around data sovereignty, AI ethics, care, and feminist research ethics within AI.

Carroll et al [42] formulate a care-centric data practice, building upon critical concerns pertaining to data sovereignty and self-management within Indigenous communities from Oceania, the United States and Canada. They offer a set of principles—CARE[2]— to complement an existing approach to data management [3] [43]. They articulate the objectives of the CARE principles as constituting people-; purpose-; and data-centric concerns. In our analytical approach, we specifically work with these as lenses for our analysis.

Gray and Witt [25] formulate a preliminary roadmap for integrating a feminist data ethics of care framework within the field of AI. They argue that the ambiguity around mainstream understandings of AI-ethics lends itself to 'fuzzy' definitions, enabling systematic failure in responsibility which in turn implicitly reinforces gender-power imbalances. Of particular note to this paper is their focused attention to both the actors (the 'who') and the practicalities (the 'how') in bringing feminist approaches and methods as a remedy-of-sorts to the principle-to-practice gap. They frame this as 'making interventions' into the economy of machine learning. They propose 5 interventionist principles for feminist data ethics and care. These encompass: 1) diversity with regards to representation and participation in the machine learning economy; 2) critique of positionality; 3) foregrounding human(s) throughout a machine learning pipeline; 4) ensuring the implementation of accountability and transparency measures; and 5) equitable distribution of responsibility.  It is these 5 principles that we have identified as assistive lenses for our critique of *Holly+*.

Gray [44] expands upon their earlier paper with Witt, articulating perspectives on how the development and advancement of AI ethics will not see significant, positive change until all stakeholders take on the responsibility of engaging with ethical work and practices throughout the entirety of the economy of machine learning. They highlight that the current landscape is "dominated by a heteropatriarchal class of men", referring to the work by Chang in [45]. Gray underlines the burning need for people working within technology fields to radically change the existing culture of these fields, which they propose as key to "build[ing] capacity for care throughout the entire machine learning economy".

## Concerning Matters and Care-fully Troubling *Holly+*

In this section, we draw together theory and methods of speculative feminism, care-ethics, and feminist data-ethics and utilise these as critical evaluative tools in analysing the artwork *Holly+* by Holly Herndon. We proceed by first providing a brief overview of what *Holly+* is, and then delve into our speculative feminist and care-ethics informed critique.

### A few words about our approach to critique of *Holly+*

We conduct our critique from a particular point-of-access, in which we occupy a position as spectators and observers of the work. We see this as in coherence with the observable intent that Herndon wishes for their work to be experienced. Our occupancy of this stance is deliberate. We have utilised only publicly accessible information regarding the *Holly+* artwork, encompassing information regarding the general model structure, its governance, affiliated parties and Herndon's recorded statements regarding this artwork. This is so that we may critically examine what is 'visibilised' in and about the work, so that we may in turn be able to critically address the components of the work that appear 'invisibilised' [46]. Our understanding of these terms proposed by Hampton—'invisibilise' and 'visibilise'—as an active choice in what components of a system are seen versus unseen and acknowledging that these choices are (potentially) accompanied by harms. We see invisibilisation and visibilisation as core concerns in connection with Gray and Witt's 4th interventionist principle: accountability and transparency.

## About *Holly+*

Created by artist Holly Herndon, the work *Holly+* is a voice model built in collaboration with [Never Before Heard Sounds](#) (NBHS hereafter), a music studio devoted to the construction, development and deployment of AI powered tools for browser-based musical production. Structurally, it is a custom deep neural network trained upon recorded voice data (constituting singing and/or speech) of Herndon, deployed as a browser-based tool where prospective users upload an audio file (presumably of their own, or publicly sourced recorded material). The model utilises pitches and rhythms from the uploaded audio file, adding additional components from the training data provided by Herndon [47]. The browser-based platform with which one can engage with *Holly+* is presented in Figure 1 below, and accessible for engagement via [the following link](#).

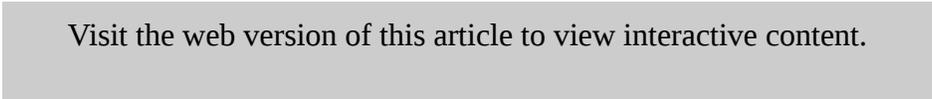

Visit the web version of this article to view interactive content.

FIGURE 1. Holly+ Browser-Based Platform

## Care-full Troubling

As a starting point for our critical evaluation, we drew inspiration from Haraway's Implosion [48] methodology as delineated by Dumit [38] to formulate a Knowledge Map and preliminary index (see Figure 2 below) of the various dimensions and structures oscillating within and around *Holly+*. We highlight, that of the 14 dimensions described by Dumit, our Knowledge Map below consists of 12 dimensions, delimited due to the scope of this paper.

| Labor dimensions | Technological dimensions; | Bodily/organic dimensions; |
|---|---|---|
| • Holly Herndon and Never Before Heard Sounds (Chris Deaner & Yotam Mann)<br>• NBHS<br>  • machine learning instruments and expressive audio tools.<br>• Financial members of the DAO<br>• Herndon's Original Dataset<br>  • The network is trained on recorded speech and singing from the target voice.<br>• Labour of DAO stewards in governing minted usage of Holly+<br>• Backend code labour<br>  • Libraries<br>  • Franeworks<br>  • API<br>• Labour of the hardware<br>  • Computers used by the coders<br>  • Computers/device used by the user | • Voice Model<br>• Deep Neural Network<br>• Raw audio generation<br>• Cloud transformation of dedicated GPU<br>• Hardware to access Holly+ platform<br>• Recording equipment used to record both Herndon's voice, and the material contributed by the platform user<br>• User added technology when implementing Holly+<br>  • ie MIDI instrument | • Inhabiting the voice of another<br>• "During my lifetime, I will exclusively retain the right to do whatever I want with my physical voice! This project exclusively concerns Holly+, my digital vocal twin 🤖"<br>  • Separation between physical self (voice) and digital twin |
| **Material dimensions;** | **Context and situated-ness;** | **Political dimensions;** |
| • Holly+<br>  • custom model on multiple hours of Holly Herndon's isolated vocal stems to create a generative instrument that retains the pitches and rhythms of a user-uploaded audio file, but adds textures and timbres learned from the training set.<br>• Computation costs<br>  • hardware<br>  • environment | • Field of voice generation<br>  • Wavenet<br>  • Tacotron<br>• Projected demand for official/high fidelity vocal models of public figures<br>• Rights to a Voice<br>  • Bette Midler court case<br>  • Tom Waits court case<br>• Music Industry<br>• Musical genre bending<br>• Sound transformation<br>  • voice transformation<br>  • instrument-instrument transformation<br>• Software - Hardware implementation<br>  • MIDI instrument application coming soon<br>• Music generation<br>  • Generating capital | • DAO and Blockchain<br>  • decision making decentralised?<br>• OpenLaw<br>  • legal agreements with Ethereum<br>    • smart contracts<br>    • drafting legal agreements<br>• Zora<br>  • media registry protocol |
| **Economic dimensions; (INCLUDE FUNDING ETC)** | **Textual dimensions;** | **Historical dimensions** |
| • DAO Stewardship<br>  • Profits from commercial usage put back into DAO<br>• This approach supports the "My Collectible Ass" principle, advocated for by ZORA and originally proposed by theorist McKenzie Wark. This principle states that the more prominent and visible/audible a work of art is, the more valuable the certified original becomes.<br>• ERC-20 Voice Tokens<br>  • distributed by Herndon (CURATORIAL)<br>• NFT minting<br>• Capital value<br>  • Capital value added to NBHS through the cultural capital of Herndon | • Smart contracts<br>  • lawyers<br>  • legal entities<br>  • court<br>• Public discourse around Holly+<br>  • Holly+ Blog<br>  • Ars Electronica<br>• Public discourse around vocal deepfakes | • Intellectual Property and Sovereignty<br>• Longitudinal recordings of Herndon's voice<br>• Voice legacy and king<br>  • Voice as a collective activity<br>    • Choirs, singing, chant<br>• History of voice-technology<br>  • voice synths<br>  • TextToSpeech generation |
| **Mythological & Symbolic Dimensions** | **Professional/Epistemological dimensions:** | **Symbolic Dimensions** |
| • Body snatching<br>• Puppetry<br>• Host and Parasite<br>• Narratives/Associations<br>  • Collective<br>  • Financial<br>  • Communist<br>  • Science<br>  • Progress | • Artists<br>  • Self management of artwork<br>  • Management of financial proceeds/profits from artwork<br>• Lawyers<br>• Commercial ML companies<br>• Media registry<br>  • copyright<br>  • royalties | • Voice control<br>• Voice distribution<br>• Human governance over machine, profiting off of it |

FIGURE 2. Knowledge Map featuring our preliminary exploration of utilising Haraway's Implosion to reveal matters of concern in *Holly+*

We then troubled the artwork through extrapolating connections extending from our central matters of concern data and identity; management and reclamation; and preservation and protection of legacy. We connected these matters of concern to principles from Carroll et al's CARE Data Principles [42], and Gray and Witt's [25] Feminist Data Ethics praxis. With regards to CARE principles, we utilised their larger categorisations of people-, purpose- and data-oriented concerns to probe how our matters of concern revealed in *Holly+* may be motivated through these larger CARE categorisations. Similarly, Gray and Witt's feminist data ethics principles were engaged as critical lenses of how matters of concern in *Holly+* may or may not be coherent with a feminist data ethics. This can be seen in Figure 3 (below), which depicts our three-layer methodological approach to analysis.

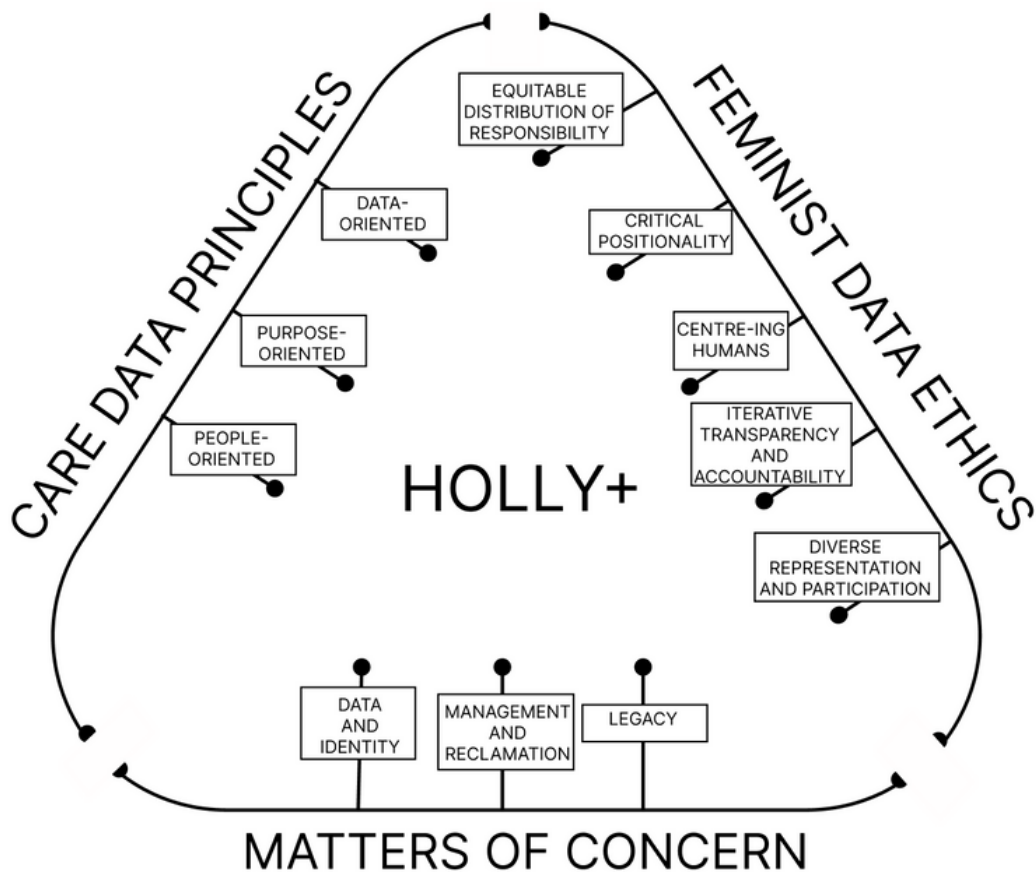

FIGURE 3. Our three-layer methodological approach to analysis, incorporating Feminist Data Ethics, Care Data Principles and Matters of Concern in *Holly+*

Our Haraway Index was highly generative in illustrating dimensions with multiple entanglements to our central matters of concern—data and identity; management and reclamation; and legacy. Of the 12 dimensions we evaluated, the most richly entangled were the technological; labour; political; and economic dimensions. We have utilised these 4 dimensions as additional Latour-informed scenography to our feminist and care-centric analysis.

## Matters of Concern in *Holly+*

From our positionality outlined above, we identify three main pillars of concern pertaining to and within *Holly+*, concerning data and identity; management and reclamation; and legacy[4].

### From Data and Identity to Management and Reclamation

One especially notable aspect of *Holly+* is the novel approach Herndon adopts in the management of artistic work engaging with the voice model as a generative tool. *Holly+* is a publicly accessible tool, and Herndon motivates her decision in open-access as an intention " …to decentralize access, decision making and profits made from my digital twin, Holly+ …" [47]. Here, we wish to cast a critical gaze over the particularities of how principles and modus operandi of speculative feminism and care ethics may (and may not) be embedded in the procedures, presentation and adjacent framings of this artwork.

It is clearly disclosed on Herndon's personal webpage that a Decentralised Autonomous Organisation (DAO) [49] stewards artistic work that deploys *Holly+*. For contextual grounding with respect to how we proceed with our critique in light of the DAO[5] stewardship, Herndon has previously engaged in discussion around decentralisation within AI-arts [50], and the reclamation of ownership of one's (literal) voice in an age of increasing concern of ethical implications of vocal deepfakes and voice synthesis [51] [52] [53] [54] [55].

They argue that the distribution of tools such as those offered through *Holly+* is in alignment with values pertaining to communality and commonality of voice. Further, they argue for DAO as a means to enable ethical, officially sanctioned and informed experimentation with another's vocal likeness and further enabling communal financial benefit in economic proceeds generated from the use of a voice model. We, however, argue that the deployment of the DAO is in fact not substantially decentralising decision-making. We argue that significant decision-making which has implications for how *Holly+* has been made and can be used, has clearly already been established by Herndon and NBHS in their design of the system, the means of interacting with *Holly+*, and the terms of agreement within the stewardship itself. The potential responsibility of stewards is thus delimited to governing 'fair-usage' of minted artworks created with *Holly+*, and not affording governance of the evolution of the architecture of *Holly+* over time. We therefore do not see the full scope of decision-making pertaining to *Holly+* as *fully* decentralised.

### Legacy

Herndon describes one of their underlying motivations for the birth of *Holly+* as an act of futuring and "maintaining the value and reputation of [their] voice [rather] than the rights being passed down to someone less familiar with the values and standards associated with [their] work". Their justification for this is grounded in concerns that inherited rights—through a next-of-kin or other Western-centric inheritance tradition—offer less posthumous protections than a public and digital distribution of governance. We do not critique Herndon's expression of feeling more comfort in distributed ownership of her voice model, we do however note interesting and "sticky" concepts entangled with this pertaining to the matter of the public following of *Holly+* and the DAO stewardship.

The first "sticky" concept we wish to highlight is the entrance procedure of the DAO [56] stewardship. Herndon outlines how membership into the DAO is contingent on the distribution of ERC-20 VOICE tokens [57] which are on the Ethereum blockchain [58]. These tokens represent voting shares in *Holly+ DAO*. These tokens will be "airdropped to collectors of my art, friends and family of the project, and other artists selected to participate in using the *Holly+* voice to create new works." We can therefore plainly assume, that *Holly+DAO* stewards, either already have a vested financial interest in Herndon's work (in the example of collectors), are already intimately familiar with Herndon and their work (friends and family of the project), or have been deemed by Herndon as possessing sufficient technical competency or musicality to create 'suitable enough' artwork using *Holly+* (other artists selected to participate in using the *Holly+* voice). We argue that the procedure for becoming a DAO stewards is highly selective, curatorial and holds the potential for exclusion based on cultural capital, digital accessibility and economic status.

This leads into our second concern, the preservation of legacy. The formation of culture does not take place in a vacuum, and there are (potentially) deeper issues in anticipating that one's values and standards may be preserved for the future production of artwork taking place in a future environment and context that we cannot yet imagine. How might the *Holly+DAO* stewards in 100-years' time be best suited or situated to make decisions that honour Herndon when living memory of Herndon as an artist may no longer exist? This assumption can be further troubled by speculating how applicable or relevant the cultural values or artistic standards of an artist may be in this selfsame future context. Stickiness and murkiness reside in the question around what constitutes an appropriate, or artistically relevant, usage of Herndon's voice especially when voting stewards may approve an offensive or uncharacteristic deployment of *Holly+* [59]. The premise of *Holly+* as an artwork is grounded—and indeed, dependent—in public interaction. The greater the engagement levels, the greater the social value that is attributed to the artwork. The DAO incentivisation scheme concretises this 'value of attention', distributing profits of artworks made with *Holly+* amongst stewards to encourage their decision-making in 'minting' usages of the voice model to increase the social capital (i.e. the visibility and distribution) of *Holly+*. We speculate on the potential stickiness of this in regards to Herndon's intention to preserve their artistic legacy. This proves especially troublesome a notion, especially when we acknowledge that the economic profits generated by any usage [60] may indeed subvert Herndon's own vision for fiscally incentivising DAO stewards to preserve her artistic legacy. Money talks, controversy sells, and the distinction between 'acceptable' and profitable are not necessarily kept apart [61].

## The visibilised and the invisibilised

Our Haraway Implosion Index further revealed imbalances pertaining to what was visibilised and invisibilised in the artwork, which we wish to draw attention to. The first is the involvement with NBHS, which we understand as having been involved in the design and development of *Holly+*. Adjacent to this invisibilised element is the code, which has not been made accessible anywhere that we could locate. Presumably, any open-source access to the code has been dismissed by the immediate partners in *Holly+* (which we speculate encompasses Herndon and NBHS) in the protection NBHS's commercial interests as an organisation developing online AI music tools. Second, we note also that there is ambiguity as to the sourcing of the original dataset of Herndon's voice, and whether this dataset was compiled specifically *for* the training of the *Holly+* model or constitutes Herndon's historical vocal data.

## CARE-ing values and *Holly+:* People-, purpose- and data-centric values

We turn now to our critical examination of people-, purpose- and data-centric values in *Holly+* in relation to our matters of concern. The larger structural design of *Holly+* as an artwork reflects values around collective interaction with identity, through the capacity to re-realise an audio recording through the consensual usage of Herndon's vocal likeness. Here, we understand that *Holly+* is prioritising both people- and purpose-centric values: by serving as a model for establishing working processes for consensual and collective engagement with identity play [62]. We understand this concentration of collective engagement and explorative play with *Holly+* as likewise reflecting principles of care- in 'demystifying' AI-tools through open-access play, and through Herndon's attention to how future demand of voice models must be informed by a system of governance that is in the best interests of the voice-origin.

Further, the legal and financial structures that govern the verified usage of *Holly+* reflect a concern towards the people-, purpose- and data-centric values proposed by Carroll et al. Herndon's vocal likeness is established as connected with their personhood[6] and image as an artist- and therefore Herndon as implicated in any future utilisation of *Holly+* in an artistic work. The DAO stewardship system appears to address these values by protecting the verified usage Herndon's vocal likeness and artistic legacy whilst enabling collective participation in the formation of musical subcultures that would utilise the usage of an artist's voice in a posthumous context [63]. We can understand this as establishing a prioritisation towards people (the stakeholders in *Holly+*; purpose (Herndon's view of future voice model usage); and data (Herndon's voice and artistic legacy as data).

However, when we further position the *Holly+* DAO in relation to matters of fact and concern, we observe a conflict of people-versus-purpose centric values. From a matters of fact position, Herndon is reclaiming ownership (of data; of voice; of their likeness) with NBHS, and enabling open-source access to their identity play. However, a matter of concern is that NBHS (and Herndon) are making very rigid decisions about how the model interacts with user-contributed material; the selectivity of what vocal material is added to the dataset the model is trained on; and how the model's verified usage may be distributed (through deploying a *Holly+DAO)*. Here is where the conflict lies: the conflict between the open-access intentionality of Herndon, with the closed-system development of the *Holly+* architecture.

## Feminist Data Ethics and *Holly+*

When we take additional lenses from Gray and Witt's 5 feminist data ethics principles, we can further understand that the matters of concern in *Holly+* become somewhat more tremulous, or ambiguous. By this, we mean that our matters of concern pertaining to *Holly+* are troubled when examined through the 5 feminist principles formulated by Gray and Witt. This therefore requires our care and attention.

With regard to the first principle—equitable distribution of responsibility—although the structure of governance of *Holly+* through the DAO aims to equitably divide decision-making power, we do not see this distribution as *truly* equitable. As had previously been discussed, the distribution of stewardship tokens is contingent on either a financial and/or labour investment in Herndon's artistic work; a familial or network (by this we also presume a cultural capital) connection to Herndon; or bestowment of a token based on Herndon's assessment of the recipient's artistic merit or capacity. We argue that this limits access, by requiring technical capacity and familiarity with—and a secure financial position to invest in—cryptotech. Based on these grounds it can be argued that these pathways to stewardship are in fact not entirely equitable.

The second principle—critical positionality—is more clearly addressed. Herndon has continually articulated their views on the future usage of voice modeling, and the inclusion of Web3 technologies to safeguard the legal interests of artists. In the third principle—the centering of human(s) throughout the pipeline—this becomes more difficult to discern. Naturally there is a centering of Herndon's capacity to share their likeness and encourage collective and creative usage of their identity play. However, we were unable to ascertain specifically how a human-centered approach was applied with regards to data collection or system design. This invites further speculation as to how the valuation of collective play—an apparent concern addressed thematically in *Holly+*—may be more transparently addressed in the design of the model's architecture and its interactivity.

The fourth principle—transparency and accountability— proves similarly more problematic for us to assess. We observed a lack of transparency with regards to the particularities of the vocal model, and specifically to the availability of the code. We assume this as withheld due to the commercial interests of Herndon and NBHS, yet this withholding necessitates clarification. It must not go unacknowledged that *Holly+* is an artwork made in collaboration with 2 entities, one an artist and one an organisation, both with vested interests in preserving certain aspects of code as their intellectual property and holding cultural and financial value.

When we further consider transparency and accountability, it is also not clear how this is factored in with regards to the *Holly+DAO*. On one hand, there is transparency with regards to how the stewardship is implemented to govern verified usages of *Holly+*. On the other hand, we were unable to find any information regarding *who* specifically had been awarded DAO stewardship. This is an area of concern, as the transparency of this system is selectively visibilised and invisibilised- and contrary to Gray and Witt's proposed principle. There is also ambiguity as to the nature of accountability in regards to actions taken by DAO stewards, present and future. We had previously ruminated on potential implications of stewards having to make decisions regarding usage of Herndon's vocal likeness in a speculative future context with—potentially—markedly different values around culture than we have in our present reality.

In the fifth and final principle—diverse representation and participation— this is perhaps the most ambiguous to determine. On the development end, *Holly+* utilises libraries and frameworks which have presumably been constructed by a particular demographic [45], with shared or complimentary technical skillsets. With regards to user engagement, its browser-hosting enables widespread, international participation. However, access is contingent on computer or smartphone access and a reliable internet connection- which can be significant factors of exclusion.

# Burning Concerns and Speculative Future Directions towards a Feminist Musical AI

Our critique reveals that AI-systems carry many residues: unintentional and intentional imprints from the datasets they have been trained on[7]; the algorithms they have been made with[8]; and the actors who then inherit or use these systems in varying applications. We see a critical need for these residues to be address, and we propose interdisciplinary feminist methods as a means to do so. This constitutes work in 3 categories: de-centralisation of management; preservation and protection of legacy; and the critical prioritisation of people-, process- or data-oriented principles.

With regards to de-centralisation, we see potential in exploring alternative structures to put 'power' back in the hands of people accessing and making the artworks, rather than a board of directors from a recording label [64]. We foresee this matter of concern as being a crucial area for future troubling of existing power structures within the music industry.

In terms of preservation and protection of legacy, we have seen the deployment of Web3 legal and financial technologies troubling current modus operandi protecting artistic legacy. Through engagement with novel systems of artwork stewardship, we foresee future 'unsettling of the waters' with how artists can protect or distribute ownership and management of their data[9] and artistic legacy. We foresee such engagement with AI technologies eliciting profound changes in the preservation and protection of the legacy of an artist.

A final[10] matter of concern is the critical prioritisation of people-, process- or data-oriented principles. We have seen these navigated in *Holly+* through the concern for the future of ethical voice model utilisation (people- and process-oriented) and subsequent implementation of DAO governance. We propose 2 central questions that researchers may utilise as a preliminary step in their implementation of feminist and care-full methods in musical-AI design: '*who and/or what is invisibilised?*'; and conversely, '*who and/or what is visibilised?*"

# Future Work

This paper is an initial peek into implementations of feminism and care ethics into musical-AI. The scope of the matters of concern addressed in this paper is substantial, with important future work needed with regards to further analysis required of hegemonic power structures within the field of musical-AI, and evaluation of how barriers of access—linguistic, social and digital—are implemented throughout the economy of AI. We further anticipate that the presentation of practical examples of conscious engagement with matters of fact, concern and care will form the basis of our future work in this regard.

# Conclusion

Within this paper, we have opened and stepped into a critical space within which we have troubled the waters of Musical-AI. We have outlined existing research work that engages with the implementation of feminist discourse, perspectives and methodologies across disciplines such as science and technology studies (STS), care ethics, and AI and Data Ethics. We have taken up a critical feminist lens of a musical-AI artwork—*Holly+*—and care-fully troubled the various dimensions within this artwork that we see as collective matters of concern across intersecting disciplines negotiating tensions around AI. Through our interdisciplinary analytical approach, we have revealed matters of concern which pose future troubling of power structures within the music industry. Further resulting from our preliminary critique through a speculative feminist lens, we address the burning matters of concern within Musical-AI and outline potential directions for future work in troubling inherited tools, systems, methodologies and lore around artistic and musical AI use.

# Ethics Statement and Acknowledgements

This work was partially supported by the Wallenberg AI, Autonomous Systems and Software Program – Humanities and Society (WASP-HS) funded by the Marianne and Marcus Wallenberg Foundation and the Marcus and Amalia Wallenberg Foundation.

# Footnotes

1. The hyphenation is intentional here, and throughout ↩
2. Collective benefit; Authority to control; Responsibility; Ethics ↩
3. The 'FAIR Guiding Principles for scientific data management and stewardship' ↩
4. As indicated in Figure 3 ↩
5. A DAO is a community-led entity that governs decision-making processes of a product or project its operating protocols are formulated into a smart contract which is written onto the blockchain. DAO members receive a distribution of profits resulting from the usage of the product , proportional to their financial investment. ↩
6. By this we do not mean personhood in the biological sense, nor a quality of the voice (such as timbre). Instead, we are referencing Herndon's own separation of her vocal self into 2: one occupying her own physical body, and the other (Holly+) as a digital vocal twin. Both are Holly, but occupying different forms. ↩
7. Which further encompasses the unintentional and intentional leakages of values, assumptions and intentionalities of both the subject whose data has been utilised and the intentionalities of the data collectors ↩

8. Encompassing the unintentional and intentional leakages of values, assumptions and intentionalities of the developers and makers of said algorithms ↩

9. In this case, their artworks ↩

10. In the context of this paper at least. ↩

## References

1. Hampton, L. M. (2021). Black Feminist Musings on Algorithmic Oppression. *Proceedings of the 2021 ACM Conference on Fairness, Accountability, and Transparency*, 1. New York, NY, USA: Association for Computing Machinery. https://doi.org/10.1145/3442188.3445929 ↩

2. Buolamwini, J., & Gebru, T. (2018). Gender Shades: Intersectional Accuracy Disparities in Commercial Gender Classification. *Proceedings of the 1st Conference on Fairness, Accountability and Transparency*, 77–91. PMLR. Retrieved from https://proceedings.mlr.press/v81/buolamwini18a.html ↩

3. Hicks, M. (2019). Hacking the Cis-tem. *IEEE Annals of the History of Computing*, *41*(1), 20–33. https://doi.org/10.1109/MAHC.2019.2897667 ↩

4. Raji, I. D., Gebru, T., Mitchell, M., Buolamwini, J., Lee, J., & Denton, E. (2020). Saving Face: Investigating the Ethical Concerns of Facial Recognition Auditing. *Proceedings of the AAAI/ACM Conference on AI, Ethics, and Society*, 145–151. New York, NY, USA: Association for Computing Machinery. https://doi.org/10.1145/3375627.3375820 ↩

5. Offert, F., & Phan, T. (2022). *A Sign That Spells: DALL-E 2, Invisual Images and The Racial Politics of Feature Space*. arXiv. https://doi.org/10.48550/arXiv.2211.06323 ↩

6. Birhane, A., Prabhu, V. U., & Kahembwe, E. (2021). *Multimodal datasets: misogyny, pornography, and malignant stereotypes*. arXiv. https://doi.org/10.48550/arXiv.2110.01963 ↩

7. Schwartz. (2018). "The discourse is unhinged": how the media gets AI alarmingly wrong. *The Guardian*. Retrieved from https://www.theguardian.com/technology/2018/jul/25/ai-artificial-intelligence-social-media-bots-wrong ↩

8. *Proposal for a REGULATION OF THE EUROPEAN PARLIAMENT AND OF THE COUNCIL LAYING DOWN HARMONISED RULES ON ARTIFICIAL INTELLIGENCE (ARTIFICIAL INTELLIGENCE ACT) AND AMENDING CERTAIN UNION LEGISLATIVE ACTS*. (2021). Retrieved from https://eur-lex.europa.eu/legal-content/EN/TXT/?uri=CELEX%253A52021PC0206 ↩

9. *Proposal for a REGULATION OF THE EUROPEAN PARLIAMENT AND OF THE COUNCIL on European data governance (Data Governance Act)*. (2020). Retrieved from https://eur-lex.europa.eu/legal-

content/EN/TXT/?uri=CELEX%253A52020PC0767 ↩

10. Carnovalini, F., & Rodà, A. (2020). Computational Creativity and Music Generation Systems: An Introduction to the State of the Art. *Frontiers in Artificial Intelligence*, *3*, 14. https://doi.org/10.3389/frai.2020.00014 ↩

11. Döbereiner, L. (2022). Artistic Potentials of Fallacies in AI Research. *Proceedings of the 3rd Conference on AI Music Creativity, AIMC.*, 5. https://doi.org/10.5281/zenodo.7088311 ↩

12. Trump, S. (2021). Musical Cyborgs: Human-Machine Contact Spaces for Creative Musical Interaction. *Proceedings of the 2nd Conference on AI Music Creativity*, 11. ↩

13. Dahlstedt, P. (2021). Musicking with Algorithms: Thoughts on Artificial Intelligence, Creativity, and Agency. In E. R. Miranda (Ed.), *Handbook of Artificial Intelligence for Music: Foundations, Advanced Approaches, and Developments for Creativity* (pp. 873–914). Cham: Springer International Publishing. https://doi.org/10.1007/978-3-030-72116-9_31 ↩

14. Born, G., & Devine, K. (2016). Gender, Creativity and Education in Digital Musics and Sound Art. *Contemporary Music Review*, *35*(1), 1–20. https://doi.org/10.1080/07494467.2016.1177255 ↩

15. Tatar, K., Ericson P., Cotton K., Núñez del Prado P. T., Batlle-Roca R. Cabrero-Daniel, B, Ljungblad S., Diapoulis G., Hussain J. *A Shift In Culture through Artificial Intelligence*. In press. ↩

16. Sturm, B., Monaghan, O., Collins, Ú., Et, D., Year, A., Sturm, B., … Pachet, F. (2018). Machine Learning Research that Matters for Music Creation: A Case Study. *Journal of New Music Research*, *In Press*. https://doi.org/10.1080/09298215.2018.1515233 ↩

17. Born, G., Morris, J., Diaz, F., & Anderson, A. (2021). *Artificial intelligence, music recommendation, and the curation of culture: A white paper* (p. 27) [Techreport]. University of Toronto; Schwartz Reisman Institute for Technology. ↩

18. Dignum, V., Casey, D., Cerratto-Pargman, T., Dignum, F., Fantasia, V., Formark, B., … Tucker, J. (2023). *On the importance of AI research beyond disciplines*. arXiv. https://doi.org/10.48550/ARXIV.2302.06655 ↩

19. Latour, B. (2014). *What Is the Style of Matters of Concern?* https://doi.org/10.5749/minnesota/9780816679959.003.0004 ↩

20. **Tatar, K.**, Ericson P., Cotton K., Núñez del Prado P. T., Batlle-Roca R. Cabrero-Daniel, B, Ljungblad S., Diapoulis G., Hussain J. *A Shift In Culture through Artificial Intelligence*. In press. ↩

21. Haraway, D. (1988). Situated Knowledges: The Science Question in Feminism and the Privilege of Partial Perspective. *Feminist Studies*, *14*(3), 575–599. https://doi.org/10.2307/3178066 ↩

22. Haraway, D. J. (2016). *Staying with the trouble: making kin in the Chthulucene*. Durham: Duke University Press. ↩

23. de la Bellacasa, M. P. (2011). Matters of care in technoscience: Assembling neglected things. *Social Studies of Science*, *41*(1), 85–106. https://doi.org/10.1177/0306312710380301 ↩

24. Gray, J., & Witt, A. (2021). A feminist data ethics of care for machine learning: The what, why, who and how. *First Monday*. https://doi.org/10.5210/fm.v26i12.11833 ↩

25. Dumit, J. (2014). Writing the Implosion: Teaching the World One Thing at a Time. *Cultural Anthropology*, *29*(2), 344–362. https://doi.org/10.14506/ca29.2.09 ↩

26. Latour, B., & Weibel, P. (2005). *Making Things Public*. Cambridge, Massachusetts: MIT Press. Retrieved from https://mitpress.mit.edu/9780262122795/making-things-public/ ↩

27. States, T. C. R. C. (2019). Monthly Review | A Black Feminist Statement. Retrieved from https://monthlyreview.org/2019/01/01/a-black-feminist-statement/ ↩

28. Coaston, J. (2019). The intersectionality wars. Retrieved from https://www.vox.com/the-highlight/2019/5/20/18542843/intersectionality-conservatism-law-race-gender-discrimination ↩

29. Ahmed, S. (2017). No [Blog]. Retrieved from https://feministkilljoys.com/2017/06/30/no/ ↩

30. Gamble, S. (2001). *The Routledge Companion to Feminism and Postfeminism*. Routledge. ↩

31. Beasley, C. (1999). *What is Feminism?: An Introduction to Feminist Theory*. SAGE Publications. Retrieved from https://libgen.li/ads.php?md5=0c10b7721de68fd4d685b49257df0ce5 ↩

32. Weedon, C. (2002). Key Issues in Postcolonial Feminism: A Western Perspective. *Gender Forum: An Internet Journal for Gender Studies*, (1). Retrieved from https://web.archive.org/web/20131203002056/http://www.genderforum.org/issues/genderealisations/key-issues-in-postcolonial-feminism-a-western-perspective/ ↩

33. Frankenberg, R. (1993). Growing up White: Feminism, Racism and the Social Geography of Childhood. *Feminist Review*, (45), 51–84. https://doi.org/10.2307/1395347 ↩

34. Henry, N., Vasil, S., & Witt, A. (2022). Digital citizenship in a global society: a feminist approach. *Feminist Media Studies*, *22*(8), 1972–1989. https://doi.org/10.1080/14680777.2021.1937269 ↩

35. Haraway, D. (2006). A Cyborg Manifesto: Science, Technology, and Socialist-Feminism in the Late 20th Century. In J. Weiss, J. Nolan, J. Hunsinger, & P. Trifonas (Eds.), *The International Handbook of Virtual Learning Environments* (pp. 117–158). Dordrecht: Springer Netherlands. https://doi.org/10.1007/978-1-4020-3803-7_4 ↩

36. Haraway, D. J. (2018). *Modest_Witness@Second_Millennium. FemaleMan_Meets_OncoMouse: Feminism and Technoscience* (2nd ed.). Second edition. | New York, NY : Routledge, 2018. | The title is an email: Routledge. https://doi.org/10.4324/9780203731093 ↩

37. Dumit, J. (2014). Writing the Implosion: Teaching the World One Thing at a Time. *Cultural Anthropology*, *29*(2), 344–362. https://doi.org/10.14506/ca29.2.09 ↩

38. Bardzell, S. (2010). Feminist HCI: taking stock and outlining an agenda for design. *Proceedings of the SIGCHI Conference on Human Factors in Computing Systems*, 1301–1310. New York, NY, USA: Association for Computing Machinery. https://doi.org/10.1145/1753326.1753521 ↩

39. Bardzell, S., & Bardzell, J. (2011). Towards a feminist HCI methodology: social science, feminism, and HCI. *Proceedings of the SIGCHI Conference on Human Factors in Computing Systems*, 675–684. New York, NY, USA: Association for Computing Machinery. https://doi.org/10.1145/1978942.1979041 ↩

40. Michelfelder, D. P., Wellner, G., & Wiltse, H. (2017). *Designing differently : toward a methodology for an ethics of feminist technology design*. Rowman & Littlefield International. Retrieved from http://urn.kb.se/resolve?urn=urn:nbn:se:umu:diva-134366 ↩

41. Carroll, S., Garba, I., Figueroa-Rodríguez, O., Holbrook, J., Lovett, R., Materechera, S., … Hudson, M. (2020). The CARE Principles for Indigenous Data Governance. *Data Science Journal*, (19), 1–12. https://doi.org/https://doi.org/10.5334/dsj-2020-042 ↩

42. Wilkinson, M. D., Dumontier, M., Aalbersberg, Ij. J., Appleton, G., Axton, M., Baak, A., … Mons, B. (2016). The FAIR Guiding Principles for scientific data management and stewardship. *Scientific Data*, *3*(1), 160018. https://doi.org/10.1038/sdata.2016.18 ↩

43. Gray, J. E. (2022). What can feminism do for AI ethics? Retrieved from https://medium.com/mlearning-ai/what-can-feminism-do-for-ai-ethics-b7e401889441 ↩

44. Chang, E. (2018). *Brotopia: Breaking Up the Boys' Club of Silicon Valley*. Portfolio. Retrieved from https://www.amazon.com.au/Brotopia-Breaking-Boys-Silicon-Valley/dp/0735213534 ↩

45. Hampton, L. M. (2021). Black Feminist Musings on Algorithmic Oppression. *Proceedings of the 2021 ACM Conference on Fairness, Accountability, and Transparency*, 1. New York, NY, USA: Association for Computing Machinery. https://doi.org/10.1145/3442188.3445929 ↩

46. Herndon, H. (2021). Holly+ 􏰀 􏰀 􏰀. Retrieved from https://holly.mirror.xyz/54ds2IiOnvthjGFkokFCoaI4EabytH9xjAYy1irHy94 ↩

47. Haraway, D. J. (2018). *Modest_Witness@Second_Millennium. FemaleMan_Meets_OncoMouse: Feminism and Technoscience* (2nd ed.). Second edition. | New York, NY : Routledge, 2018. | The title is an email: Routledge. https://doi.org/10.4324/9780203731093 ↩

48. Shuttleworth, D. (2021). What Is A DAO And How Do They Work? Retrieved from https://consensys.net/blog/blockchain-explained/what-is-a-dao-and-how-do-they-work/ ↩

49. Minsker, E. (2021). Holly Herndon's AI Deepfake "Twin" Holly+ Transforms Any Song Into a Holly Herndon Song. Retrieved from https://pitchfork.com/news/holly-herndons-ai-deepfake-twin-holly-transforms-any-song-into-a-holly-herndon-song/ ↩

50. Khanjani, Z., Watson, G., & Janeja, V. (2021). *How Deep Are the Fakes? Focusing on Audio Deepfake: A Survey*. ↩

51. Cheng, H., Guo, Y., Wang, T., Li, Q., Chang, X., & Nie, L. (2022). *Voice-Face Homogeneity Tells Deepfake*. ↩

52. Coldewey, D. (2023). VALL-E's quickie voice deepfakes should worry you, if you weren't worried already. Retrieved from https://techcrunch.com/2023/01/12/vall-es-quickie-voice-deepfakes-should-worry-you-if-you-werent-worried-already/ ↩

53. Almutairi, Z., & Elgibreen, H. (2022). A Review of Modern Audio Deepfake Detection Methods: Challenges and Future Directions. *Algorithms*, *15*, 19. https://doi.org/10.3390/a15050155 ↩

54. Khanjani, Z., Watson, G., & Janeja, V. (2023). Audio deepfakes: A survey. *Frontiers in Big Data*, *5*, 1001063. https://doi.org/10.3389/fdata.2022.1001063 ↩

55. Decentralized autonomous organizations (DAOs). (2023). Retrieved from https://ethereum.org ↩

56. What Are ERC-20 Tokens on the Ethereum Network? (n.d.). Retrieved from https://www.investopedia.com/news/what-erc20-and-what-does-it-mean-ethereum/ ↩

57. What is Ethereum? (n.d.). Retrieved from https://ethereum.org ↩

58. Ars-Electronica. (2022). Holly+. Retrieved from https://starts-prize.aec.at/en/holly-plus/ ↩

59. Egan, M. (2020). The art world has a money laundering problem | CNN Business. Retrieved from https://www.cnn.com/2020/07/29/business/art-money-laundering-sanctions-senate/index.html ↩

60. Hahn, L. (2020). Artists and their controversies: a question of profit? Retrieved from https://artgateblog.altervista.org/artists-and-their-controversies-a-question-of-profit/ ↩

61. Live, U. (2022). *IP for the AI Era: Holly+ Presents Identity Play*. Unfinished Live 2022. Retrieved from https://live.unfinished.com/videolibrary/mainstage-i-ip-for-the-ai-era-holly-presents-ident ↩

62. Mbembé, J.-A., & Meintjes, L. (2003). Necropolitics. *Public Culture*, *15*, 11–40. ↩

63. Brook, T. (2023). Music, Art, Machine Learning, and Standardization. *Leonardo*, *56*(1), 81–86. https://doi.org/10.1162/leon_a_02135 ↩